\documentclass[aps,twocolumn,superscriptaddress,showpacs]{revtex4}
\usepackage{graphicx}
\usepackage{dcolumn}
\usepackage{bm}
\usepackage{enumerate}
\begin{document}
\title{Threshold configurations 
in the presence of Lorentz violating dispersion relations
}
\author{D. Mattingly}
\email[]{davemm@physics.umd.edu}
\author{Ted Jacobson}
\email[]{jacobson@physics.umd.edu}
\author{S. Liberati}
\email[]{liberati@physics.umd.edu}
\affiliation{Physics Department, University of Maryland, College
Park, MD 20742-4111, USA}
\date{8 October 2002; file {\sf threshpaper.tex} --- version 0.1;
\LaTeX-ed \today}
\bigskip
\begin{abstract}
\bigskip

A general characterization of lower and upper threshold configurations
for two particle reactions is determined under the assumptions that
the single particle dispersion relations $E(|{\bf p}|)$ are
rotationally invariant and monotonic in $|{\bf p}|$, and that energy
and momentum are conserved and additive for multiple particles. It is
found that at a threshold the final particle momenta are always
parallel and the initial momenta are always anti-parallel. The
occurrence of new phenomena not occurring in a Lorentz invariant
setting, such as upper thresholds and asymmetric pair production
thresholds, is explained, and an illustrative example is given.

\end{abstract}
\maketitle
\input epsf
\def\wt{\widetilde}
\def\gsim{\; \raisebox{-.8ex}{$\stackrel{\textstyle >}{\sim}$}\;}
\def\lsim{\; \raisebox{-.8ex}{$\stackrel{\textstyle <}{\sim}$}\;}
\def\half{{1\over2}}
\def\a{\alpha}
\def\b{\beta}
\def\g{\gamma}
\def\d{\delta}
\def\e{\epsilon}
\def\o{\omega}
\def\m{\mu}
\def\L{{\mathcal L}}
\def\d{{\mathrm{d}}}
\def\p{{\mathbf{p}}}
\def\q{{\mathbf{q}}}
\def\k{{\mathbf{k}}}
\def\fp{{p_{\rm 4}}}
\def\fq{{q_{\rm 4}}}
\def\fk{{k_{\rm 4}}}
\def\etal{{\emph{et al}}}
\def\det{{\mathrm{det}}}
\def\tr{{\mathrm{tr}}}
\def\ie{{\emph{i.e.}}}
\def\aka{{\emph{aka}}}
\def\R{{\cal R}}
\def\HRULE{{\bigskip\hrule\bigskip}}

\section{Introduction}

The possibility of Lorentz violation (LV) is currently receiving much
attention~\cite{CPT}. One of the reasons for this is that high energy
observations are now capable of detecting Planck suppressed
LV~\cite{Amelino-Camelia:1997gz,CG,ACP,Kluzniak,Kifune:1999ex}, which
could be a harbinger of quantum gravity
effects~\cite{KS89,loopqg,Amelino-Camelia:2001cm}.  A straightforward way to
probe LV is though the consequences of Lorentz violating dispersion
relations for particles.  Kinematics with LV can allow new reactions,
some with thresholds, or can modify thresholds for reactions.  LV
threshold effects are of particular interest because they are
generically sensitive to Planck suppressed LV at energies far below
the Planck energy allowing the use of current astrophysical
observations for obtaining good constraints on LV
parameters~\cite{CG,Aloisio:2000cm,JLM01,Major,JLM02}.

The purpose of this paper is to establish some general properties of
threshold configurations which are needed for studying reactions in
the presence of Lorentz violation. These properties have been used
without proof in most of the previous literature on the subject (the
only exception being Ref.~\cite{CG}). We prove a {\it threshold
configuration theorem} which asserts that for arbitrary rotationally
invariant, monotonically increasing dispersion relations the final
particle momenta of a two particle reaction are always parallel and
the initial momenta are always anti-parallel at a threshold.  The
proof of this theorem involves first establishing that, at any (upper
or lower) threshold, the configuration of initial and final particles
is such as to minimize the energy of the final particles subject to
momentum conservation at fixed values of the incoming particle
energies.  Our results are a generalization of those found in
Ref.~\cite{CG}, which addressed just particle decays with quadratic
dispersion relations allowing for a different maximum speed for each
particle type. After proving the threshold theorem, the occurrence of
new phenomena not occurring in a Lorentz invariant setting, such as
upper thresholds and asymmetric pair production, is explained, and an
illustrative example is given.

\section{Threshold configuration theorem}

We consider reactions with two initial and two final particles.
Results for reactions with only one incoming or outgoing particle can
be obtained as special cases.  We allow each particle to have an
independent dispersion relation $E({\bf p})$, where $\bf p$ is the
three momentum and $E$ is the energy, and make the following
\begin{quotation}
 \noindent{\bf Assumptions}:
 \noindent \begin{enumerate}
 \item $E({\bf p})$ is a rotation-invariant function of $\bf p$.
 \item $E({\bf p})$ is a monotonic increasing function of $|{\bf p}|$.
 \item Energy and momentum are both additive for multiple particles and
conserved.
 \end{enumerate}
\end{quotation}
{\it Comments on the assumptions}: 1) If rotation invariance is not
assumed the question of threshold relations is obviously much more
complicated. 2) The monotonic assumption is necessary; without it the
threshold theorem would not be true. Note however that the dispersion
need only be monotonic in the observationally relevant range of $|{\bf
p}|$. 3) The third assumption follows from the assumption of
space-time translation invariance, with $E$ and ${\bf p}$ interpreted
as the usual generators of time and space translations. Assumptions 1)
and 3) are by no means inevitable, and indeed there is much research
on Lorentz violation that does not make some of these assumptions (see
e.g.~\cite{Kc,Ellis,Mersini}).

\begin{figure}[htb]
\vbox{ \vskip 10 pt
\centerline{\epsfxsize=3in\epsffile{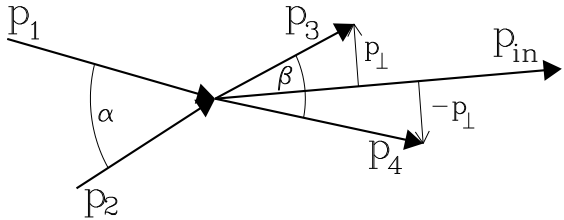}}
\caption{Geometry of a two particle
reaction.~\label{fig:4part}.} \smallskip}
\end{figure}

A number of variables are needed to describe the kinematics of a two
particle reaction of the form in Fig.~\ref{fig:4part}.
We denote the magnitude of the 3-momenta of the two incoming particles
by $p_1$ and $p_2$, and the angle between them by $\alpha$. The final
particle momentum magnitudes are $p_3$ and $p_4$. The angle between
the total incoming momentum vector ${\bf p}_{\rm in}={\bf p}_1+{\bf
p}_2$ and one of the outgoing momenta ${\bf p}_3$ is $\beta$. Since
the dispersion relations are assumed rotationally invariant there is
no loss of generality in assuming all of the momenta lie in a single
plane.

A {\it threshold} is defined {\it relative to a fixed value of} $p_2$
as follows. With $p_2$ fixed, we ask for what values of $p_1$ is there
an energy and momentum conserving configuration
$(\alpha,\beta,p_3,p_4)$? If there exists a $p_L$ such that when
$p_1<p_L$ there are no allowed configurations then we call $p_L$ a
{\it lower threshold}.  Likewise if there exists a $p_U$ such that
when $p_1>p_U$ there are no allowed configurations then we call $p_U$
an {\it upper threshold}. To prove the threshold configuration theorem
it is helpful to understand the solution space of the conservation
equations in a graphical manner as follows.

\begin{figure}[htb]
\vbox{ \vskip 10 pt
\centerline{\epsfxsize=2.7in\epsffile{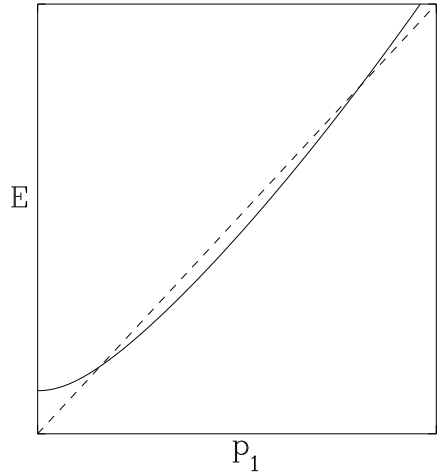}}
\caption{\label{fig:Ef} $E_f^{\alpha,\beta,p_3} (p_1)$ The dashed line
is a null line given for reference.
\smallskip}}
\end{figure}

For given values of $(p_1,p_2,\alpha,\beta,p_3)$ momentum conservation
determines $p_4$ and therefore also the energy of the final particles:
\begin{equation}
E_f(p_1,p_2,\alpha,\beta,p_3)=E_3(p_3) + E_4(p_4(p_1,p_2,\alpha,\beta,p_3)).
\end{equation}
For each value of the configuration variables  $(p_2,\alpha,\beta, p_3)$
we can thus define the final energy function:
\begin{equation}
E_f^{\alpha,\beta,p_3}(p_1) =E_f(p_1,p_2,\alpha,\beta,p_3).
\end{equation}
Since $p_2$ is fixed in the definition of a given threshold, that
label is suppressed in the notation $E_f^{\alpha,\beta,p_3}$. A
plot of $E_f^{\alpha,\beta,p_3}(p_1)$ looks for example like Fig.
\ref{fig:Ef} (assuming that $E_f^{\alpha,\beta,p_3}(0)>0$).

Any choice of $\alpha,\beta,p_3$ gives a curve $E_f^{\alpha,\beta,p_3}$
that is non-negative since each individual particle's dispersion relation is
non-negative. Now consider the region $\cal R$ 
in the $E,p_1$ plane covered
by plotting $E_f^{\alpha,\beta,p_3}$ for all possible
configurations $\alpha,\beta,p_3$. Since each
of these curves is bounded
below there exists a lower
boundary $E_B$ to this region, determined by the configuration of
lowest final energy at each $p_1$. 
\begin{figure}[htb]
\vbox{ \vskip 10 pt
\centerline{\epsfxsize=2.7in\epsffile{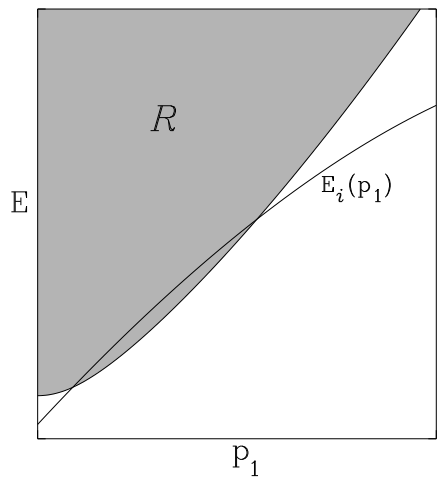}}
\caption{\label{fig:Efall} $\R$  is the region covered by all
curves $E_f^{\alpha,\beta,p_3} (p_1)$ for some fixed $p_2$,
assuming momentum conservation holds to determine $p_4$. The curve
$E_i(p_1)$ is the initial energy for the same fixed $p_2$. Where
the latter curve lies inside  $\R$  there is a solution to the
energy and momentum conservation equations.
\smallskip}
}
\end{figure}

The initial energy $E_i(p_1,p_2)$ is just a function of $p_1$ for
fixed $p_2$, so it can be plotted on the same graph. An example
is shown in Fig. \ref{fig:Efall}. In the figure it is assumed
that the reaction does not happen for $p_1=0$, i.e. that
$E_f^{\alpha,\beta,p_3}(0)>E_i(0)$ for all choices of
$\alpha,\beta,p_3$. If $E_f^{\alpha,\beta,p_3}(0)\leq E_i(0)$
then all of the following results still apply, but there is
simply no lower threshold.

The plot in Fig. \ref{fig:Efall} is a graphical representation of
the solution space to the conservation equation. There exists a
solution when $E_i$ is in  $\R$ and no solution when $E_i$ is
outside of $\R$.  It is evident  that a lower threshold occurs
when $E_i$ enters  $\R$ for the first time and an upper threshold
occurs when it leaves  $\R$ for the last time. The threshold
momenta $p_L$ and $p_U$ therefore correspond to the intersections
of $E_i$ with $E_B$. Hence we have
the following important result:
\begin{quotation}
\noindent  {\bf Minimum energy theorem:} 
\emph{An upper or lower threshold configuration occurring at
incoming momentum $p_1$ is always  the minimum energy
configuration conserving momentum at that $p_1$.}
\end{quotation}

Using this result  we can now establish the
\begin{quotation}
\noindent  {\bf Threshold configuration theorem:} 
\emph{At an upper or lower threshold the incoming particle momenta
are always anti-parallel and the final particle momenta are parallel.}
\end{quotation}
That is, at a threshold, the
angles are necessarily given by $\beta=0$ and $\alpha=\pi$. To 
prove the first statement, note that 
any value of $\beta$ other than 0 or $\pi$, momentum conservation
can be preserved while simultaneously decreasing $p_3$ and $p_4$,
since one can 
 subtract out from ${\bf p}_3$ and ${\bf p}_4$
equal and opposite momenta ${\bf p}_\perp$ and $-{\bf p}_\perp$
transverse to ${\bf p}_{\rm in}$ (see Fig. \ref{fig:4part}. For
$\beta =\pi$,  ${\bf p}_3$ and ${\bf p}_4$ must be anti-parallel,
hence momentum conservation can again be preserved while
decreasing both their magnitudes. Since all dispersion relations
are assumed monotonic, any operation that decreases $p_3$ and $p_4$ 
also decreases the energies $E_3$ and
$E_4$ of each of the final particles and so decreases
$E_f^{\alpha,\beta,p_3}(p_1)$. Only $\beta=0$ can therefore be
the lowest energy configuration, hence at a threshold the final
particle momenta must always be parallel.

As to the initial particles, if $\alpha<\pi$ then at fixed
$p_1,p_2$,   the total incoming momentum can be lowered  by
increasing $\alpha$. Since the final particles are necessarily
parallel to each other and to the total momentum, monotonicity
implies that this allows $E_f$ to be lowered, e.g. by reducing
$p_4$ at fixed $\beta=0$ and $p_3$. Therefore $\alpha<\pi$ cannot
be a threshold configuration, hence the initial particle momenta
must always be anti-parallel at a threshold.

As an aside, note that $E_i$ can leave and enter  $\R$ multiple
times.  These intermediate intersection points look like
thresholds locally in $p_1$ but are not global thresholds.
Nevertheless they could in principle yield interesting
phenomenology. Since they do not occur for simple models of
Lorentz violating dispersion however, we will not analyze them in
detail here. We simply mention that their kinematic configuration
will be the same as for global thresholds and their locations in
$p_1$ determined in the same manner.

\section{New threshold phenomena from Lorentz violation}
\subsection{Upper thresholds}

The possibility of upper thresholds in the presence of Lorentz
violating dispersion was noticed in Ref.~\cite{CG} for particle decays
and in Ref.~\cite{Kluzniak} was discussed in the context of photon
annihilation to an electron-positron pair $\gamma\gamma \rightarrow
e^+ e^-$.  In~\cite{JLM02} the analysis was generalized to allow for
the possibility of asymmetric pair production (discussed in the next
subsection) and a wider class of dispersion relations. Here we make
some general remarks about the the origin of this peculiar feature of
Lorentz violating kinematics.

If Lorentz invariance holds, then there is never an upper threshold.
This is can be seen by transforming to the center of mass frame, in
which the total incoming four-momentum is $(E,0,0,0)$.  Suppose the
final particles have masses $m_1$ and $m_2$, and that $E\ge
m_1+m_2$. Then energy and momentum are conserved with the final
particle four-momenta $(E_3(|{\bf p}|),{\bf p})$ and $(E_4(|{\bf
p}|),-{\bf p})$ for some choice of ${\bf p}$. As $p_1$ grows, the
condition $E\ge m_1+m_2$ is always met for a head-on collision with
Lorentz invariant dispersion, hence there is no upper threshold.

This argument exploited the Lorentz invariant form of the dispersion
relations to guarantee that 1) the total four-momentum is timelike, 2)
the energy of each particle in the center of mass frame is an
unbounded positive function of the momentum $p$, and 3) the center of
mass energy increases without bound as $p_1$ increases. Any of these
properties can fail to hold for non Lorentz invariant dispersion
relations. In particular, if a monotonic unbounded Lorentz violating
dispersion relation is transformed from the preferred frame to another
frame, it will in general no longer be monotonic or unbounded.

That upper thresholds exist for LV dispersion relations is evident
from Fig.~\ref{fig:Efall}. Since different particles can have
different dispersion, the curve $E_i$ can be varied independently from
$E_f$ and hence $E_B$ by choosing various dispersion relations for the
initial particles (assuming the initial and final particles are
different).  If we choose a dispersion relation such that $E_i$ enters
and leaves the region $\R$ then this reaction has an upper and lower
threshold for that choice of dispersion.

\subsection{Asymmetric pair production}

Unlike in Lorentz invariant physics, reactions with identical final
particle dispersion can have  (both upper and lower) 
threshold configurations for which the two final momenta are not equal.
This was first noticed in Ref.~\cite{JLM01} in the context of 
photon decay and photon annihilation. The example of photon
decay will be discussed in the next section.
We start by analyzing the Lorentz invariant case and then
show how it is modified when Lorentz invariance is violated.

A familiar result in Lorentz invariant physics is that the threshold
configuration for pair production of massive particles is symmetric in
the final momentum distribution, i.e. $p_3=p_4$. This follows
immediately from the fact that in the center of mass frame the final
particles must be created at rest at threshold, so in any other frame
they have equal momenta (since they have the same mass).  To
understand the Lorentz violating case it is useful to re-derive this
fact without using the Lorentz transformation to the center of mass
frame, since the dispersion relations are not Lorentz invariant and
one cannot even always boost to the center of mass frame.

At any threshold the final momenta are parallel and the initial
momenta are anti-parallel, and momentum conservation
holds. Suppose that $p_3 > p_4$.  Momentum
conservation can be preserved while while lowering $p_3$ and raising
$p_4$ by an equal amount.  In a Lorentz invariant theory the
dispersion relation $E_o(p_3)$ for each of the outgoing (massive)
particles always has positive curvature with respect to $p_3$.
Therefore $\partial E_o/\partial p$ is greater at $p=p_3$ than at
$p=p_4$. Since $p_3$ and $p_4$ are changed by opposite amounts we have
\begin{equation}
\Delta E_f=\left. -\Delta p \frac {\partial E_o} {\partial p}
\right|_{p=p_3}
+ \left. \Delta p \frac {\partial E_o} {\partial p} \right|_{p=p_4}<0
\end{equation}
Therefore $E_f$ can be lowered, implying that $p_3>p_4$ is not a
threshold.  Similarly $p_4$ cannot be greater than $p_3$, hence
$p_3=p_4$ at a threshold.

The previous argument does not depend specifically on Lorentz
invariance.  Hence it establishes a general result:
\begin{quotation}
\noindent{\bf Condition for symmetry of pair production}:
\emph{Pair production is always symmetric at threshold provided
the outgoing particle dispersion relation has positive curvature with 
respect to momentum.}
\end{quotation}
However,  in the Lorentz violating case $E_o(p)$ need not  
have positive curvature with respect to $p$, 
in which case asymmetric pair production is possible.
At an asymmetric threshold, the above argument shows that
$E_o(p)$ must have negative curvature.

A sufficient condition for the threshold configuration to be {\it not}
symmetric can be found by computing the variation $\Delta
E_f=\Delta(E_3+E_4)$ induced by variations $\Delta p_3=-\Delta p$ and
$\Delta p_4=+\Delta p$ away from the symmetric solution $p_3=p_4$.
The first order variations of $E_3$ and $E_4$ cancel, so we have
\begin{eqnarray}
\Delta E_f&=&\left[
\frac{1}{2}\frac {\partial^2 E_o}{\partial p^2}(-\Delta p)^2 +
\frac{1}{2} \frac {\partial^2 E_o}{\partial p^2}(\Delta p)^2\right]_{p=p_3}
 \nonumber \\&=&
\left. \frac {\partial^2 E_o} {\partial p^2}
\right|_{p=p_3}\, (\Delta p)^2.
\end{eqnarray}
Thus $E_f$ can be lowered by moving away from the $p_3=p_4$
configuration if the dispersion relations are such that $\partial^2
E_o/\partial p^2<0$ at $p=(p_1-p_2)/2$, where $E_i(p_1)$ intersects
the boundary of $\R$. In such a case the outgoing particle momenta are
not symmetric at the threshold. This condition on $\partial^2
E_o/\partial p^2$ is sufficient but not necessary for having an
asymmetric threshold, since the energy may be locally minimized by the
symmetric configuration but not globally minimized.

\section{Example: $\gamma\rightarrow e^+e^-$}

To illustrate the general results we now discuss the case of photon
decay to an electron-positron pair.  The example is chosen for its
simplicity, so one can easily see how the threshold phenomena depend
on the values of the Lorentz violating parameters. We refer to
\cite{JLM02} for analysis of further cases of direct observational
relevance.

Consider the following deformed photon and electron 
dispersion relations respectively:
\begin{eqnarray}
\o^2&=& k^2 + \epsilon k^2 + \xi k^3/M \\ 
E^2 &=& p^2 + m^2 + \eta p^3/M
\end{eqnarray}
where $\epsilon$, $\xi$ and $\eta$ are dimensionless parameters and
$M=10^{19}$ GeV is approximately the Planck mass.  The physical idea
is that these are just the deformations of lowest order in $p/M$.
Observed photons have energies far below $M$, so we will calculate the
threshold in the regime where $p/M \ll 1$.  Hereafter we adopt units
with $M=1$.

In the Lorentz-invariant case $\epsilon=\xi=\eta=0$ the photon cannot
decay. The final energy region $\R$ asymptotes to the line $E_f=p$ in
the $E$-$p$ plane, while the incoming energy curve is just
$E_i(p)=\omega(p)=p$ which never enters $\R$. The simplest way to
allow photon decay is to let $\epsilon$ or $\xi$ be positive and
nonzero. In this case, $E_i(p)$ enters $\R$ and never leaves, so there
is a lower threshold but no upper threshold. Since the
electron/positron energy function $E(p)$ has positive curvature, this
lower threshold configuration is symmetric in the final momenta.

To allow for an upper threshold we can take $\epsilon>0$ and $\xi<0$
(with $\eta=0$). At sufficiently large $k$ the $k^3$ term in the
photon dispersion relation will cause the $E_i(k)$ to dip back down
and exit $\R$, again at a symmetric configuration.  To determine the
value of $k$ at these thresholds we invoke the threshold configuration
theorem and the positive curvature of $E(p)$ to set the final momenta
each equal to $k/2$. Energy conservation then yields
\begin{equation}
\xi k^3 +\epsilon k^2 = 4m^2. 
\label{eq:caseA}
\end{equation} 
The lower and upper thresholds occur at the lower and upper positive
roots of Eq. (\ref{eq:caseA}).  For a numerical example, let us
suppose that $\xi=-1$.  The maximum of the left hand side of
(\ref{eq:caseA}) occurs at $k=(2/3)\epsilon$, where it is equal to
$k^3/2= (4/27)\epsilon^3$.  In order to have a threshold this must be
larger than $4m^2$, so $\epsilon\ge3m^{2/3}= 4.1\times 10^{-15}$.  The
threshold is at $k = 2m^{2/3} = 28$ TeV for this critical value of
epsilon, and for larger epsilon the lower threshold is below 28 TeV
while the upper threshold is above 28 TeV.  

To obtain an example with asymmetric thresholds we must introduce
negative curvature in the electron dispersion relation, hence we allow
for $\eta<0$. The case with $\epsilon=0$ and $\xi\ne0$ was studied in
detail in Ref.~\cite{JLM02}, where it was found that there is never an
upper threshold and there can be a lower threshold which is either
symmetric or asymmetric, depending on the values of $\xi$ and $\eta$.
We now sketch how that conclusion is reached.

Since the final momenta are not necessarily equal at threshold, the
algebra is more complicated.  To keep it manageable we expand the
dispersion relations to lowest order in the small quantities. Using
the fact that the final momenta are parallel at threshold we set the
positron momentum equal to $k-p$, and the energy conservation equation
then becomes
\begin{equation}
\xi k^2 = \eta (p^2+(k-p)^2) + \frac {m^2 k} { p (k-p)},
\label{eq:econ}
\end{equation}
This equation is valid as long as both $p$ and $k-p$ are much greater
than the mass $m$, and all momenta are much smaller than $M$. (For a
careful discussion about the validity of the expansion see
Ref.~\cite{JLM02}.

Were we to incorrectly assume the threshold to be always symmetric, we
would set $p=k/2$ in Eq. (\ref{eq:econ}) and conclude that it is
given by
\begin{equation}
k^{\rm sym}_{\rm th}=
\displaystyle{ \left(\frac{8 m^2}{2\xi-\eta} \right)^{1/3}}
\label{eq:sym}
\end{equation}
This would also imply that there is only a threshold when
$\xi>\eta/2$. However, this result is incorrect.  A threshold occurs
when $E_f$ is minimized for a fixed $k$.  While the value $p=k/2$
always corresponds to a stationary point of $E_f$ (since the final
particles have equal mass), this point can be a maximum rather than a
minimum. (In general it could also be a local minimum that is not the
global minimum, but that possibility does not occur in the present
example.) For $\eta>0$, the electron dispersion relation has positive
curvature, so the symmetric point is the minimum and the symmetric
threshold (\ref{eq:sym}) applies.  When $\eta$ is negative, as shown
in Ref.~\cite{JLM02}, the symmetric threshold applies for $\xi>0$, but
for $\xi<0$ there is an asymmetric threshold whenever $\xi>\eta$:
\begin{equation}
   k^{\rm aysm}_{\rm th}=
\displaystyle{\left[\frac{-8 m^2\eta}{\left(\xi-\eta \right)^2}
\right]^{1/3}}
\label{eq:asym}
\end{equation}
The amount of asymmetry is given by $p-(k/2)=k\sqrt{\xi/\eta}$.

If one incorrectly assumed that the threshold were always symmetric,
one would miss the region $\eta<\xi<\eta/2<0$ where there is an
asymmetric threshold and no symmetric solution. Moreover, in the
overlapping region $\eta/2<\xi<0$ where the actual threshold is
asymmetric but a symmetric solution does exist, the symmetric
threshold formula (\ref{eq:sym}) would simply give the wrong result
for the threshold. For a numerical example, let $\eta=-1$.  Then for
$\xi=-0.6$ there is an asymmetric threshold $k^{\rm aysm}_{\rm
th}=50$ TeV and no symmetric solution.  For $\xi=-0.4$ there is again
an asymmetric threshold $k^{\rm aysm}_{\rm th}=38$ TeV, and a
(non-threshold) symmetric solution $k^{\rm sym}_{\rm th}=46$ TeV.

\section{Conclusion}

Much of the apparent simplicity of the familiar structure of
thresholds in Lorentz invariant theories derives from the fact that
one can always transform to the center of mass frame.  We have seen
that in a Lorentz violating context there can be surprising threshold
behavior. Nevertheless, as we have shown, certain basic kinematic
relations always pertain to threshold configurations as long as the
dispersion relations are rotationally invariant and monotonically
increasing with momentum.  These relations can be used to
systematically determine the threshold behavior.
 

\end{document}